\documentclass[journal,onecolumn, draftcls, 12pt]{IEEEtran}

\usepackage{url}
\usepackage[normalem]{ulem}
	
\usepackage{graphicx} % for pdf, bitmapped graphics files
\usepackage{epsfig} % for postscript graphics files
\usepackage{epstopdf}
\usepackage{amssymb}  % assumes amsmath package installed
\usepackage{verbatim}
\usepackage{multirow} % For multi rows in tables
\usepackage{dsfont}
\usepackage{verbatim}
\usepackage{subcaption}
\usepackage{bbm}
\usepackage{wrapfig}
\usepackage{amsmath}
\newtheorem{theorem}{Theorem}%  meant for continuous numbers
\newtheorem{assumption}{Assumption}
\newtheorem{lemma}{Lemma}
\newtheorem{proposition}{Proposition}
\newtheorem{definition}{Definition}%
	
\newcommand{\mA}{\mathcal{A}}
\newcommand{\mX}{\mathcal{X}}

\newcommand{\mN}{[N]}
\newcommand{\cP}{\mathcal{P}}

\newcommand{\cS}{\mathcal{S}}
\renewcommand{\hat}{\widehat}

\def\lpr{\big\{}
\def\rpr{\big\}}

\newcommand{\mE}{\mathbb{E}}	

\newcommand{\eq}[1]{\begin{align}#1\end{align}}
\newcommand{\seqn}[1]{\begin{subequations}#1\end{subequations}}
\newcommand{\lb}[1]{\big\{ \begin{array}{ll} #1 \end{array} }

\newcommand{\E}{\mathbb{E}}
 \newcommand{\nn}{\nonumber}
\newcommand{\cX}{\mathcal{X}}
\newcommand{\cZ}{\mathcal{Z}}

\newcommand{\cA}{\mathcal{A}}

\newcommand{\tsigma}{\tilde{\sigma}}
\newcommand{\hz}{\hat{z}}

\newcommand{\cH}{\mathcal{H}}
\newcommand{\tgamma}{\tilde{\gamma}}

\newcommand{\defeq}{\buildrel\triangle\over =}

\newcommand{\pushright}[1]{\ifmeasuring@ #1 \else\omit\hfill$\displaystyle#1$\fi\ignorespaces}
\newcommand{\pushleft}[1]{\ifmeasuring@ #1 \else\omit$\displaystyle#1$\hfill\fi\ignorespaces}

\newcommand\bbx{\mathbf{x}}
\newcommand\bbz{\mathbf{z}}

\newcommand\calE{\mathcal{E}}
\newcommand\calA{\mathcal{A}}
\newcommand\calP{\mathcal{P}}

\newcommand\strat{\mathcal{S}}

\begin{document}
%%%%%%%%%%%%%%%%

%\setcounter{page}{1}
%\DOI{10.1287/xxxx.0000.0000}%


%\RUNTITLE{Mean field teams and games with correlated types}

% \TITLE{Bundling Information Goods of Decreasing Value}
\title{Mean field teams and games with correlated types}
 %% Single author, or several authors with same affiliation:
 \author{%
  \IEEEauthorblockN{Deepanshu Vasal}\\
  \IEEEauthorblockA{Northwestern University, Evanston, IL, USA\\
  dvasal@umich.edu
 }
 }

%\KEYWORDS{deterministic inventory theory; infinite linear programming duality;
%  existence of optimal policies; semi-Markov decision process; cyclic schedule}

%\KEYWORDS{Game Theory, Mean field games, Sequential decomposition, Backward recursion}

\maketitle
%%%%%%%%%%%%%%%%%%%%%%%%%%%%%%%%%%%%%%%%%%%%%%%%%%%%%%%%%%%%%%%%%%%%%%

% Samples of sectioning (and labeling) in MNSC
% NOTE: (1) \section and \subsection do NOT end with a period
%       (2) \subsubsection and lower need end punctuation
%       (3) capitalization is as shown (title style).
%
%\section{Introduction.}\label{intro} %%1.
%\subsection{Duality and the Classical EOQ Problem.}\label{class-EOQ} %% 1.1.
%\subsection{Outline.}\label{outline1} %% 1.2.
%\subsubsection{Cyclic Schedules for the General Deterministic SMDP.}
%  \label{cyclic-schedules} %% 1.2.1
%\section{Problem Description.}\label{problemdescription} %% 2.

% Text of your paper here

\begin{abstract}
     Mean field games have traditionally been defined~\cite{LaLi07,HuMaCa06} as a model of large scale interaction of players where each player has a private type that is independent across the players.
In this paper, we introduce a new model of mean field teams and games with \emph{correlated types} where there are a large population of homogeneous players sequentially making strategic decisions and each player is affected by other players through an aggregate population state. Each player has a private type that only she observes and types of any $N$ players are correlated through a kernel $Q$. All players commonly observe a correlated mean-field population state which represents the empirical distribution of any $N$ players' correlated joint types. We define the Mean-Field Team optimal Strategies (MFTO) as strategies of the players that maximize total expected joint reward of the players. We also define Mean-Field Equilibrium (MFE) in such games as solution of coupled Bellman dynamic programming backward equation and Fokker Planck forward equation of the correlated mean field state, where a player's strategy in an MFE depends on both, her private type and current correlated mean field population state. We present sufficient conditions for the existence of such an equilibria. We also present a backward recursive methodology equivalent of master's equation to compute all MFTO and MFEs of the team and game respectively. Each step in this methodology consists of solving an optimization problem for the team problem and a fixed-point equation for the game. We provide sufficient conditions that guarantee existence of this fixed-point equation for the game for each time $t$.
% Enter your abstract
\end{abstract}%

\section{Introduction}
To model the behavior of large population strategic interactions, Mean Field Games (MFGs) were introduced independently by \cite{LaLi07} and \cite{HuMaCa06}. In such games, there are a large number of homogeneous strategic players, each with an independent type that evolves as a controlled Markov process, where each player has infinitesimal effect on system dynamics and is affected by other players through a mean-field population state. Since its introduction, there have been a large number of applications such as economic growth, security in networks, oil production, volatility formation, population dynamics (see ~\cite{La08,GuLaLi11,SuMa19,HuMa16,HUMa17,HuMa17cdc, AdJoWe15} and references therein). 

However, many time in the real world the players have \emph{correlated} preferences. For instance, when people are voting for a candidate, or buying a product, or are getting infected by a malware, their state can be correlated across players, for example, your voting preferences may be correlated with that of your neighbors, whose preferences maybe correlated with their neighbors and so on. Such a scenario with correlated types can not be captured by the traditional MFG models presented in~\cite{HuMaCa06,LaLi07}, which significantly reduces the applicability of such models.

In this paper, we introduce a new model that corrects this shortcoming. We consider discrete-time mean field teams and games where each player sequentially makes strategic decisions and is affected by other players through a \emph{correlated} mean-field population state. Each player has a private type that evolves through a controlled Markov process which only she observes and all players observe the current population state which is the distribution of other players' types. We assume that types of the players are correlated in the society such that for any $N$ player, their types evolve through a known symmetric kernel. 

In the team version of the problem, players have an objective to maximize the total expected common reward of the players. In the corresponding games, when each agent has a homogeneous reward function that only depends on her own type, action and the common correlated mean field state, a Mean Field Equilibrium (MFE) is defined through a coupled backward-forward equation as follows: the correlated mean-field state evolves through Fokker Planck \emph{forward} equation given an MFE policy profile of the players. And MFE policy satisfies the Bellman \emph{backward} equation, given the correlated mean-field states. As a result, in order to compute an MFE, one needs to solve a coupled backward and forward fixed-point equation in the space of correlated mean-field states and equilibrium policies.
%MFE is shown to be a good approximation of Nash equilibrium (or MPE) of the original game as the number of players grow large (for instance see ~\cite{Caetal15,La16,Fi17,La18,DeLaRa19} and references therein).

%
In this paper, we consider a non-stationary model where players are \emph{cognizant} i.e. they actively observe the current population state (which need not have converged) and act based on that population state and their own private state. For the team problem, we provide a backward recursive dynamic program to compute optimum homogeneous Markovian strategies of the players within the class of those strategies.
For the game problem, we provide a backward recursive methodology to compute all (non-stationary) MFE of that game which involves solving a smaller fixed point equation for each time $t$. Since this methodology computes all MFE of the game, there exists a solution to this smaller fixed point equation for each time $t$, whenever there exists an MFE. Our methodology is motivated by the developments in the theory of dynamic games with asymmetric information in~\cite{VaSiAn16arxiv, VaAn16allerton, VaAn16cdc, OuTaTe17,Ta17, VaBe20}, where authors in these works have considered different models of such games and provided a sequential decomposition framework to compute Markovian perfect Bayesian equilibria or Mean field equilibria of such games. 

The paper is structured as follows. In Section~\ref{sec:Model}, we present model, notation and background. In section~\ref{sec:methodology_a}, we present a dynamic program to compute optimum Markovian homogeneous strategies for the finite horizon and infinite horizon team problem. In section~\ref{sec:methodology}, we present a methodology to compute MFE for the finite horizon game. In Section~\ref{sec:ih}, we extend the sequential decomposition idea to infinite horizon games. In Section~\ref{sec:exists}, we discuss the existence of per time fixed-point equation. We conclude in Section~\ref{sec:Concl}.
\subsection{Notation}
We use uppercase letters for random variables and lowercase for their realizations. For any variable, subscripts represent time indices and superscripts represent player identities. We use notation $-i$ to represent all players other than player $i$ i.e. $ -i = \{1,2, \ldots i-1, i+1, \ldots, N \}$. We use notation $a_{t:t'}$ to represent the vector $(a_t, a_{t+1}, \ldots a_{t'})$ when $t'\geq t$ or an empty vector if $t'< t$. We use $a_t^{-i}$ to mean $(a^1_t, a^2_{t}, \ldots, a_t^{-1}, a_t^{+1} \ldots, a^N_{t})$ . We remove superscripts or subscripts if we want to represent the whole vector, for example $a_t$  represents $(a_t^1,\ldots, a_t^N)$. We denote the indicator function of any set $A$ by $\mathbbm{1}\{A\}$. 
For any finite set $\mathcal{S},\mathcal{P}(\mathcal{S})$ represents the space of probability measures on $\mathcal{S}$ and $\vert\mathcal{S}\vert$ represents its cardinality. We denote by $P^{\sigma}$ (or $E^{\sigma}$) the probability measure generated by (or expectation with respect to) strategy profile $\sigma$. We denote the set of real numbers by $\mathbb{R}$. %For a probabilistic strategy profile of players $(\sigma_t)_{i\in [N]}$ where probability of action $a_t$ conditioned on $z_{1:t},x_{1:t}$ is given by $\sigma_t(a_t\vert z_{1:t},x_{1:t})$, we use the short hand notation $\sigma_t^{-i}(a_t^{-i}\vert z_{1:t},x_{1:t}^{-i})$ to represent $\prod_{j\neq i}\sigma_t^j(a_t^j\vert z_{1:t},x_{1:t}^j)$. 
All equalities and inequalities involving random variables are to be interpreted in \emph{a.s.} sense.

\section{Model and Background}
\label{sec:Model}
We consider a discrete time large population sequential team and game as follows. There are $M$ homogeneous players, where $M$ tends to $\infty$. We denote the set of homogeneous players by $[M]$ and with some abuse of notation, set of time by [T] for both finite and infinite time horizon. In each period $t\in[T]$, player $i\in[M]$ observes a private type $x_t\in\cX = \{1,2,\cdots, N_x\}$ and a common observation $z_t\in\cZ$, takes action $a_t\in\cA = \{1,2,\cdots, N_a \}$, and receives a reward $R(x_t,a_t,z_t)$ which is a function of its current type $x_t$, action $a_t$ and the common observation $z_t$. 
%Let $g_t(a)$ be the fraction of population playing action $a\in\cA$ at time $t$. i.e.
%\eq{
%g_t(a) = \frac{1}{N}\sum_{i=1}^N \mathbbm{1}\{a_t = a\}
%} 
%Let $g_t = (g_t(1), g_t(2), \cdots g_t(N_a))$, where $\sum_{i=1}^{N_a} g_t(i) = 1$.
Any $N$ players' types evolve as a \emph{correlated} controlled Markov process,
\eq{
(x^1_{t+1},x_{t+1}^2,\ldots, x_{t+1}^N) = f_x(x^1_{t},x_{t}^2,\ldots, x_{t}^N, a^{[1:N]}_t, z_t, w_t).
}
The random variables $(W_t)_{t}$ are assumed to be mutually independent across players and across time, and independent of initial random variables $x_1,z_1$. We also write the above update of $x_t$ through a kernel, 
\eq{
(x^1_{t+1},x_{t+1}^2,\ldots, x_{t+1}^N)\sim Q_x(\cdot\vert z_t,x^1_{t},x_{t}^2,\ldots, x_{t}^N,a^1_{t},a_{t}^2,\ldots, a_{t}^N).
}
where $Q_x$ is a kernel symmetric across all agents i.e. one can use any permutation of the order of the agents without any change in the output of the kernel. %\footnote{As will be seen later, the mean field $z_t$ is a deterministic dynamic process which implies defining a set of strategies of the players define the corresponding mean field states for the whole time horizon. Thus at any times $t$, in effect, all the players know or observe the whole mean field trajectory $z_{1:t}$.} 
She takes action $a_t^i$ according to a behavioral strategy $\sigma = (\sigma_t)_t$, where $\sigma_t:(\cZ)^{t}\times\mathcal{X}^t \to \mathcal{P}(\mathcal{A})$. We denote the space of such measurable strategies as $\cS_{\sigma}$. This implies $A_t\sim \sigma_t(\cdot\vert z_{1:t},x_{1:t})$. %We denote $\mathcal{H}_t^c = \mathcal{Z}^t$ to be the space of population states $z_{1:t}$ till time $t$. 
We denote $\mathcal{H}_t=\mathcal{Z}^t \times \mathcal{X}^t$ to be the set of observed histories $(z_{1:t},x_{1:t})$ of a player.

For finite time-horizon team, $\mathbb{T}_{T}$, all players together want to maximize their total expected discounted reward over a time horizon $T$, discounted by discount factor $0<\delta\leq1$, 
\eq{
J^{Team}_{T} :=\E^{\sigma} \left[\sum_{t=1}^T \delta^{t-1} R(X_t,A_t,Z_t) \right].
} 

For finite time-horizon game, $\mathbb{G}_{T}$, each player wants to maximize its total expected discounted reward over a time horizon $T$, discounted by discount factor $0<\delta\leq1$, 
\eq{
J^{Game}_{T} :=\E^{\sigma} \left[\sum_{t=1}^T \delta^{t-1} R(X_t,A_t,Z_t) \right].
} 

Similarly we define an infinite time-horizon team and game, $\mathbb{T}_{\infty}$ and $\mathbb{G}_{\infty}$, respectively, by replacing $T$ above by $\infty$.

%For the infinite time-horizon game, $\mathbb{G}_{\infty}$, each player wants to maximize its total expected discounted reward over an infinite-time horizon discounted by discount factor $0<\delta<1$, 
%\eq{
%J^{\infty} :=\E^{\sigma} \left[\sum_{t=1}^\infty \delta^{t-1} R(X_t,A_t,Z_t) \right].
%} 
In the following, we define the appropriate solution concepts to analyze this system.

\subsection{Solution concept: Team optimal solution}
%\subsubsection{Nash equilibrium} %and Markov perfect equilibrium}
%The Nash equilibrium (NE) of $\mathbb{G}_{T}$ is defined as strategies $\tsigma = (\tsigma_t)_{i\in[N],t\in[T]}$ that satisfy, for all $i\in[N]$, 
%\eq{
%\E^{(\tsigma,\tsigma^{-i})}[\sum_{t=1}^T \delta^{t-1} R(X_t,A_t,Z_t) ]\geq \E^{(\sigma,\tsigma^{-i})}[\sum_{t=1}^T \delta^{t-1} R(X_t,A_t,Z_t)],
%} 
%For sequential games, however, a more appropriate equilibrium concept is Markov perfect equilibrium (MPE)~\cite{MaTi01}, which we use in this paper. We note that an MPE is also a Nash equilibrium of the game, although not every Nash equilibrium is an MPE.
%An MPE $(\tsigma)$ satisfies sequential rationality such that for $\mathbb{G}_T$, $\forall i\in[N], t \in [T], h_t \in \cH_t, {\sigma}$,

%\eq{
%&\E^{(\tsigma \tsigma^{-i})}[\sum_{n=t}^T \delta^{n-t} R(X_n,A_n,Z_n)\vert z_{1:t},x_{1:t} ] \geq \nn\\
%&\E^{({\sigma} \tsigma^{-i})}[\sum_{n=t}^T \delta^{n-t} R(X_n,A_n,Z_n)\vert z_{1:t},x_{1:t} ], \;\; \;\;   \label{eq:seqeq2}
%} 
%NE for $\mathbb{G}_{\infty}$ is defined in a similar way where summation in the above equation is taken such that $T$ is replaced by $\infty$.
A team optimal solution is defined as set of symmetric Markovian strategies $\tsigma=\{ \tsigma_t\}_{t\in[T]} $, where $\tsigma_t:\cZ\times\cX\to\Delta(\cA)$ i.e. $A_t\sim \sigma_t(\cdot\vert z_{t},x_{t})$, and mean field states $z=\{ z_t\}_{t\in[T]}$ that satisfy the following optimization problem 
\begin{itemize}
\item A policy $\tsigma$ is Mean Field Team Optimal (MFTO) if for all $ t\in[T], \sigma_{t:T}, z_{1:t}$,
\eq{
&\E^{\tsigma_{t:T}}[\sum_{n=t}^T \delta^{n-t} \sum_{x_n}Z_n(x_n)R(x_n,A_n,Z_n)\vert z_{1:t} ] \geq \nn\\
&\E^{{\sigma_{t:T}}}[\sum_{n=t}^T \delta^{n-t}\sum_{x_n}Z_n(x_n) R(x_n,A_n,Z_n)\vert z_{1:t} ], \;\; \;\;   \label{eq:seqeq2}
} 
% \item Define the above backward optimization equation $\psi: \cS_z\to2^{\cS_{\sigma}}$ as
%  \eq{
% \psi(z) :=\{\sigma\in S_{\sigma} : \sigma \text{ is optimal given } z\}.
% }
% \item Conversely, define a forward mapping $\Lambda: \cS_{\sigma} \to \cS_{z} $ as follows: given $\sigma \in S_{\sigma}, z = \Lambda(\sigma)$, is constructed recursively as

% \eq{
% &z_{t+1}(x^1_{t+1},x_{t+1}^2,\ldots, x_{t+1}^N) = \sum_{x_t^{[1:N]},a_t^{[1:N]}}z_t(x^1_{t},x_{t}^2,\ldots, x_{t}^N)\times\nn\\
% & Q_x(x^1_{t+1},x_{t+1}^2,\ldots, x_{t+1}^N\vert z_t,x^{[1:N]}_{t}, a_t^{[1:N]})\prod_{i=1}^N\sigma_t(a^i_t\vert z_t,x^i_t)
% }

\end{itemize}
We note that in the above equation, team optimality is defined only within the class of Markovian policies that depend on the current state $x_t$ and the mean field population state $z_t$.

MFTO for $\mathbb{T}_{\infty}$ are defined in a similar way where summation in the above equations is taken such that $T$ is replaced by $\infty$.
%
%We note that standard MFE with independent types as defined in~\cite{LaLi07,HuMaCa06} is a special case where $N=1$.

\subsection{Solution concept: Mean Field Equilibrium}
%\subsubsection{Nash equilibrium} %and Markov perfect equilibrium}
%The Nash equilibrium (NE) of $\mathbb{G}_{T}$ is defined as strategies $\tsigma = (\tsigma_t)_{i\in[N],t\in[T]}$ that satisfy, for all $i\in[N]$, 
%\eq{
%\E^{(\tsigma,\tsigma^{-i})}[\sum_{t=1}^T \delta^{t-1} R(X_t,A_t,Z_t) ]\geq \E^{(\sigma,\tsigma^{-i})}[\sum_{t=1}^T \delta^{t-1} R(X_t,A_t,Z_t)],
%} 
%For sequential games, however, a more appropriate equilibrium concept is Markov perfect equilibrium (MPE)~\cite{MaTi01}, which we use in this paper. We note that an MPE is also a Nash equilibrium of the game, although not every Nash equilibrium is an MPE.
%An MPE $(\tsigma)$ satisfies sequential rationality such that for $\mathbb{G}_T$, $\forall i\in[N], t \in [T], h_t \in \cH_t, {\sigma}$,

%\eq{
%&\E^{(\tsigma \tsigma^{-i})}[\sum_{n=t}^T \delta^{n-t} R(X_n,A_n,Z_n)\vert z_{1:t},x_{1:t} ] \geq \nn\\
%&\E^{({\sigma} \tsigma^{-i})}[\sum_{n=t}^T \delta^{n-t} R(X_n,A_n,Z_n)\vert z_{1:t},x_{1:t} ], \;\; \;\;   \label{eq:seqeq2}
%} 
%NE for $\mathbb{G}_{\infty}$ is defined in a similar way where summation in the above equation is taken such that $T$ is replaced by $\infty$.
Mean-Field Equilibrium (MFE) is defined as set of symmetric Markovian strategies $\tsigma=\{ \tsigma_t\}_{t\in[T]} $, where $\tsigma_t:\cZ\times\cX\to\Delta(\cA)$ i.e. $A_t\sim \sigma_t(\cdot\vert z_{t},x_{t})$, and mean field states $z=\{ z_t\}_{t\in[T]}$ that satisfy the forward-backward equations defined through following equations 
\begin{itemize}
\item A policy $\tsigma$ is optimal for $z := z_{1:t}$ if for all $ t\in[T], \sigma_{t:T},x_{1:t}$,
\eq{
&\E^{\tsigma_{t:T}}[\sum_{n=t}^T \delta^{n-t} R(X_n,A_n,Z_n)\vert z_{1:t},x_{1:t} ] \geq \nn\\
&\E^{{\sigma_{t:T}}}[\sum_{n=t}^T \delta^{n-t} R(X_n,A_n,Z_n)\vert z_{1:t},x_{1:t} ], \;\; \;\;   \label{eq:seqeq2}
} 
\item Define the above backward optimization equation $\psi: \cS_z\to2^{\cS_{\sigma}}$ as
 \eq{
\psi(z) :=\{\sigma\in S_{\sigma} : \sigma \text{ is optimal given } z\}.
}
\item Conversely, define a forward mapping $\Lambda: \cS_{\sigma} \to \cS_{z} $ as follows: given $\sigma \in S_{\sigma}, z = \Lambda(\sigma)$, is constructed recursively as

\eq{
&z_{t+1}(x^1_{t+1},x_{t+1}^2,\ldots, x_{t+1}^N) = \sum_{x_t^{[1:N]},a_t^{[1:N]}}z_t(x^1_{t},x_{t}^2,\ldots, x_{t}^N)\times\nn\\
& Q_x(x^1_{t+1},x_{t+1}^2,\ldots, x_{t+1}^N\vert z_t,x^{[1:N]}_{t}, a_t^{[1:N]})\prod_{i=1}^N\sigma_t(a^i_t\vert z_t,x^i_t)
}

\end{itemize}
\begin{definition}
\label{def:MFE}
A pair $(\sigma,z)$ is an MFE if $\sigma\in\psi(z)$ and $z=\Lambda(\sigma)$.
\end{definition}
MFE for $\mathbb{G}_{\infty}$ are defined in a similar way where summation in the above equations is taken such that $T$ is replaced by $\infty$.
We note that standard MFE with independent types as defined in~\cite{LaLi07,HuMaCa06} is a special case where $N=1$.

\subsection{Existence of MFE}
\label{sec:Exist_MFE}
\begin{assumption}[A1]
Suppose the reward function $R(z_t,x_t,a_t)$ is continuous in $z_t$.
\end{assumption}
We note that the above assumption implies that the reward function is bounded.

It was shown in~\cite[Theorem~1]{DoGaGa19} that an MFE exists for a game with independent types under Assumption~A1. In the following, we show that similar arguments under kernel $Q_x$ go through to show existence of MFE with the correlated types. 

\begin{proposition}\label{thm:mfe-existence}
Under Assumption~A1, there exists an MFE of the game.
\end{proposition}
\begin{IEEEproof}
Please see Appendix~\ref{app:A0}.
\end{IEEEproof}

\subsection{Common agent approach}
Similar to the common agent approach in~\cite{NaMaTe13}, an alternate and equivalent way of defining the strategies of the players is as follows. We first generate partial function $\gamma_t:\cX\to\cP(\cA)$ as a function of $z_t$ through an equilibrium generating function $\theta_t:\cZ\to(\cX\to\cP(\cA))$ such that $\gamma_t = \theta_t[z_t]$. Then action $A_t$ is generated by applying this prescription function $\gamma_t$ on player $i$'s current private information $x_t$, i.e. $A_t\sim \gamma_t(\cdot\vert x_t)$. Thus $A_t\sim \sigma_t(\cdot\vert z_{t},x_t) = \theta_t[z_t](\cdot\vert x_t)$.

For a given symmetric prescription function $\gamma_t = \theta[z_t]$, the statistical correlated mean-field $z_t$ evolves according to the discrete-time Fokker Planck equation~\cite{ArMa14}, $\forall y\in\cX$:
\eq{
&z_{t+1}(x^1_{t+1},x_{t+1}^2,\ldots, x_{t+1}^N) = \sum_{x_t^{[1:N]},a_t^{[1:N]}}z_t(x^1_{t},x_{t}^2,\ldots, x_{t}^N)\times\nn\\
& Q_x(x^1_{t+1},x_{t+1}^2,\ldots, x_{t+1}^N\vert z_t,x^{[1:N]}_{t}, a_t^{[1:N]})\prod_{i=1}^N\gamma_t(a^i_t\vert x^i_t)
}
which implies
\eq{
z_{t+1}= \phi(z_t,\gamma_t).
}

\section{A methodology to compute MFTO policies}
\label{sec:methodology_a}
In this section, we will provide a dynamic program to compute MFTO for both $\mathbb{T}_{T}$ and $\mathbb{T}_{\infty}$. This allows one to solve smaller optimization problem for each time $t$ that equivalently solves the dynamic optimization problem across time.

%We introduce a common belief $z_t(z_t)$ defined as $z_t(z_t) := P^{\sigma}(z_t\vert z_{1:t})$. 
As mentioned before, in MFTO, strategies of player $i$ which depend on the mean field population state at time $t$, $z_{t}$, and on its current type $x_t$.

%For a given symmetric function $\theta_t$, the statistical correlated mean-field $z_t$ evolves according to the discrete-time Fokker Planck equation, $\forall y\in\cX$:
%\eq{
%z_{t+1}(y) =\sum_{x\in\cX}\sum_{a\in \cA} z_t(x)\theta_t[z_t](a\vert x)Q_x(y\vert x,a,z_t), \label{eq:z_update}
%}
%which implies
%\eq{
%z_{t+1}= \phi(z_t,\gamma_t).
%}

\subsection{Dynamic program for $\mathbb{T}_{T}$} \label{sec:fhbr}
In this subsection, we will provide a dynamic programming methodology to generate symmetric Markovian MFTO strategies of $\mathbb{G}_{T}$ of the form described above.
We generate a reward-to-go function $(V_t)_{t\in[T]}$, where $V_t:\cZ\times\cX\to\mathbb{R}$.
These quantities are generated through the optimization problem as follows.
\begin{itemize}
\item[1.] Initialize $\forall z_{T+1}$,
\eq{
V_{T+1}(z_{T+1}) \defeq 0.   \label{eq:VT+1_c}
}

\item[2.] For $t = T,T-1, \ldots 1, \ \forall z_t$, let $\theta_t[z_t] $ be generated as follows. Set ${\gamma}^*_t = \theta_t[z_t]$, where ${\gamma}^*_t$ is the solution of the following optimization problem, 
  \eq{
 {\gamma}^*_t \in  \arg\max_{\gamma_t} \E^{\gamma_t} \left[ R(X_t,A_t,z_t) +\delta V_{t+1}(\phi(z_t,\gamma_t)) \vert  z_t\right] , \label{eq:m_FP_c}
  }
 where expectation in \eqref{eq:m_FP} is with respect to random variable $(X_t,A_t)$ through the measure
$z_t(x_t)\gamma_t(a_t\vert x_t)$.%in the proof of Lemma~\ref{fact:L1} and in particular Claim~\ref{claim:C1}.

Furthermore, using the quantity ${\gamma}^*_t$ found above, define
\eq{
V_{t}(z_t) \defeq & \E^{{\gamma}^*_t} \left[ R(x_t,A_t,z_t) +\delta V_{t+1}(\phi(z_t,\gamma^*_t)) \vert  z_t\right].  \label{eq:Vdef_c}
}
   \end{itemize}
   
Then, an optimum strategy is defined as 
\eq{
{\sigma}^*_t(a_t\vert z_{1:t},x_{1:t}) = {\gamma}^*_t(a_t\vert x_t), \label{eq:sigma_fh}
} 
where ${\gamma}^*_t = \theta[z_t]$. 

In the following theorem, we show that the strategy thus constructed is an MFTO strategy 

%%%%%%%%%%%%
\begin{theorem}
\label{Thm:Main}
A strategy $(\tsigma)$ constructed from the above methodology is an MFTO i.e. $\forall t\in[T], h_t \in \cH_t, {\sigma}$,
\eq{
&\E^{\tsigma}[\sum_{n=t}^T \delta^{n-t} R(X_n,A_n,Z_n)\vert z_{1:t} ] \nn\\
&\geq \E^{{\sigma}}[\sum_{n=t}^T \delta^{n-t} R(X_n,A_n,Z_n)\vert z_{1:t}], \;\; \;\;   \label{eq:prop}
} 
and $z_{n+1} = \phi(z_n,\tsigma(\cdot\vert z_n,\cdot)) \forall n\in[T]$ such that $n\geq t$.
\end{theorem}
\begin{IEEEproof}
We first note that $\{z_t,\gamma_t\}_t$ is a controlled Markov process of this system since
\eq{
z_{t+1} = \phi(z_t,\gamma_t)
}
and 
\eq{
\mathbb{E}[R(X_t,A_t,Z_t)\vert z_t]&= \sum_{x_t}z_t(x_t)\gamma_t(a_t\vert x_t)R(x_t,a_t,z_t)\nn\\
&=\hat{R}(z_t,\gamma_t)
}

Thus one can find the optimal policies of the players using the dynamic program in~\eqref{eq:VT+1_c}--\eqref{eq:Vdef_c} using standard Markov decision theory~\cite{KuVa86}.
\end{IEEEproof}

\subsection{Dynamic program for $\mathbb{T}_{\infty}$} \label{sec:fhbr}
In this subsection, we will provide a dynamic programming methodology to generate symmetric Markovian MFTO strategies of $\mathbb{G}_{T}$ of the form described above.
We generate a reward-to-go function $(V_t)_{t\in[T]}$, where $V_t:\cZ\times\cX\to\mathbb{R}$.
These quantities are generated through the optimization problem as follows.
\begin{itemize}
\item[1.] Initialize $\forall z_{T+1}$,
\eq{
V_{T+1}(z_{T+1}) \defeq 0.   \label{eq:VT+1_c}
}

\item[2.] For $t = T,T-1, \ldots 1, \ \forall z_t$, let $\theta_t[z_t] $ be generated as follows. Set ${\gamma}^*_t = \theta_t[z_t]$, where ${\gamma}^*_t$ is the solution of the following optimization problem, 
  \eq{
 {\gamma}^*_t \in  \arg\max_{\gamma_t} \E^{\gamma_t} \left[ R(X_t,A_t,z_t) +\delta V_{t+1}(\phi(z_t,\gamma_t)) \vert  z_t\right] , \label{eq:m_FP_c}
  }
 where expectation in \eqref{eq:m_FP} is with respect to random variable $(X_t,A_t)$ through the measure
$z_t(x_t)\gamma_t(a_t\vert x_t)$.%in the proof of Lemma~\ref{fact:L1} and in particular Claim~\ref{claim:C1}.

Furthermore, using the quantity ${\gamma}^*_t$ found above, define
\eq{
V_{t}(z_t) \defeq & \E^{{\gamma}^*_t} \left[ R(x_t,A_t,z_t) +\delta V_{t+1}(\phi(z_t,\gamma^*_t)) \vert  z_t\right].  \label{eq:Vdef_c}
}
   \end{itemize}
   
Then, an optimum strategy is defined as 
\eq{
{\sigma}^*_t(a_t\vert z_{1:t},x_{1:t}) = {\gamma}^*_t(a_t\vert x_t), \label{eq:sigma_fh}
} 
where ${\gamma}^*_t = \theta[z_t]$. 

%%%%%%%%%%%%
\begin{theorem}
\label{Thm:Main}
A strategy $(\tsigma)$ constructed from the above methodology is an MFTO i.e. $\forall t\in[T], h_t \in \cH_t, {\sigma}$,
\eq{
&\E^{\tsigma}[\sum_{n=t}^T \delta^{n-t} R(X_n,A_n,Z_n)\vert z_{1:t} ] \nn\\
&\geq \E^{{\sigma}}[\sum_{n=t}^T \delta^{n-t} R(X_n,A_n,Z_n)\vert z_{1:t}], \;\; \;\;   \label{eq:prop}
} 
and $z_{n+1} = \phi(z_n,\tsigma(\cdot\vert z_n,\cdot)) \forall n\in[T]$ such that $n\geq t$.
\end{theorem}
\begin{IEEEproof}
We first note that $\{z_t,\gamma_t\}_t$ is a controlled Markov process of this system since
\eq{
z_{t+1} = \phi(z_t,\gamma_t)
}
and 
\eq{
\mathbb{E}[R(X_t,A_t,Z_t)\vert z_t]&= \sum_{x_t}z_t(x_t)\gamma_t(a_t\vert x_t)R(x_t,a_t,z_t)\nn\\
&=\hat{R}(z_t,\gamma_t)
}

Thus one can find the optimal policies of the players using the dynamic program in~\eqref{eq:VT+1_c}--\eqref{eq:Vdef_c} using standard Markov decision theory~\cite{KuVa86}.
\end{IEEEproof}

\section{A methodology to compute MFE }
\label{sec:methodology}
We first note that in the definition of MFE in Definition~1, $\sigma$ and $z$ are coupled through a fixed point equation defined through a backward equation $\psi$ and a forward equation $\Lambda$. This is a fixed point equation \emph{across time} whose complexity increases exponentially with time, and thus suffers from the same curse of dimensionality as any dynamic optimization problem.

In this section, we will provide a backward recursive sequential decomposition methodology to compute MFE for both $\mathbb{G}_{T}$ and $\mathbb{G}_{\infty}$. This allows one to solve smaller fixed -point equations for each time $t$ that equivalently solves this bigger fixed point equation across time (and is thus equivalent to dynamic program for a dynamic optimization problem where one can solve for the bigger optimization across time by solving for smaller optimization problem for each time $t$).
%We introduce a common belief $z_t(z_t)$ defined as $z_t(z_t) := P^{\sigma}(z_t\vert z_{1:t})$. 
As mentioned before, in MFE, strategies of player $i$ which depend on the mean field population state at time $t$, $z_{t}$, and on its current type $x_t$.\footnote{Note however, that the unilateral deviations of the player are considered in the space of all strategies.} Equivalently, player $i$ takes action of the form $A_t\sim \sigma_t(\cdot\vert z_t,x_t)$. %Similar to the common agent approach in~\cite{NaMaTe13}, an alternate and equivalent way of defining the strategies of the players is as follows. We first generate partial function $\gamma_t:\cX\to\cP(\cA)$ as a function of $z_t$ through an equilibrium generating function $\theta_t:\cZ\to(\cX\to\cP(\cA))$ such that $\gamma_t = \theta_t[z_t]$. Then action $A_t$ is generated by applying this prescription function $\gamma_t$ on player $i$'s current private information $x_t$, i.e. $A_t\sim \gamma_t(\cdot\vert x_t)$. Thus $A_t\sim \sigma_t(\cdot\vert z_{t},x_t) = \theta_t[z_t](\cdot\vert x_t)$.

We are only interested in symmetric equilibria of such games such that $A_t\sim \gamma_t(\cdot\vert x_t) = \theta_t[z_t](\cdot\vert x_t)$ i.e. there is no dependence of the identity of the players on their strategies.

% For a given symmetric prescription function $\gamma_t = \theta[z_t]$, the statistical correlated mean-field $z_t$ evolves according to the discrete-time Fokker Planck equation~\cite{ArMa14}, $\forall y\in\cX$:
% \eq{
% &z_{t+1}(x^1_{t+1},x_{t+1}^2,\ldots, x_{t+1}^N) = \sum_{x_t^{[1:N]},a_t^{[1:N]}}z_t(x^1_{t},x_{t}^2,\ldots, x_{t}^N)\times\nn\\
% & Q_x(x^1_{t+1},x_{t+1}^2,\ldots, x_{t+1}^N\vert z_t,x^{[1:N]}_{t}, a_t^{[1:N]})\prod_{i=1}^N\gamma_t(a^i_t\vert x^i_t)
% }
% which implies
% \eq{
% z_{t+1}= \phi(z_t,\gamma_t).
% }
%For a given symmetric function $\theta_t$, the statistical correlated mean-field $z_t$ evolves according to the discrete-time Fokker Planck equation, $\forall y\in\cX$:
%\eq{
%z_{t+1}(y) =\sum_{x\in\cX}\sum_{a\in \cA} z_t(x)\theta_t[z_t](a\vert x)Q_x(y\vert x,a,z_t), \label{eq:z_update}
%}
%which implies
%\eq{
%z_{t+1}= \phi(z_t,\gamma_t).
%}

\subsection{Backward recursive methodology for $\mathbb{G}_{T}$} \label{sec:fhbr}
In this subsection, we will provide a methodology to generate symmetric MFE of $\mathbb{G}_{T}$ of the form described above.
We define an equilibrium generating function $(\theta_t)_{t\in[T]}$, where $\theta_t:\cZ\to\{\cX\to\mathcal{P}(\cA) \}$, where for each $z_t $, we generate $\tgamma_t = \theta_t[z_t]$. In addition, we generate a reward-to-go function $(V_t)_{t\in[T]}$, where $V_t:\cZ\times\cX\to\mathbb{R}$.
These quantities are generated through a fixed-point equation as follows.
\begin{itemize}
\item[1.] Initialize $\forall z_{T+1}, x_{T+1}\in \cX$,
\eq{
V_{T+1}(z_{T+1},x_{T+1}) \defeq 0.   \label{eq:VT+1}
}

\item[2.] For $t = T,T-1, \ldots 1, \ \forall z_t$, let $\theta_t[z_t] $ be generated as follows. Set $\tilde{\gamma}_t = \theta_t[z_t]$, where $\tilde{\gamma}_t$ is the solution of the following fixed-point equation\footnote{We discuss the existence of solution of this fixed-point equation in Section~\ref{sec:exists}.}, $\forall i \in [N],x_t\in \cX$,
  \eq{
 \tilde{\gamma}_t(\cdot\vert x_t) \in  \arg\max_{\gamma_t(\cdot\vert x_t)} \E^{\gamma_t(\cdot\vert x_t)} \left[ R(X_t,A_t,z_t) +\delta V_{t+1}(\phi(z_t,\tgamma_t), X_{t+1}) \vert  z_t,x_t\right] , \label{eq:m_FP}
  }
 where expectation in \eqref{eq:m_FP} is with respect to random variable $(X_t^{[1:N-1]},A_t,X_{t+1})$ through the measure
$z_t(x_t^{[1:N-1]})\prod_{i=1}^{N-1}\gamma_t(a_t^i\vert x_t^i)Q_x(x_{t+1}\vert z_t,x_t^{[1:N-1]},x_t,a^{[1:N-1]}_t,a_t)$. %in the proof of Lemma~\ref{fact:L1} and in particular Claim~\ref{claim:C1}.
We note that the solution of~\eqref{eq:m_FP}, $\tgamma_t$, appears both on the left of~\eqref{eq:m_FP} and on the right side in the update of $z_t$, and is thus unlike the fixed-point equation found in Bayesian Nash equilibrium.

Furthermore, using the quantity $\tilde{\gamma}_t$ found above, define
\eq{
V_{t}(z_t,x_t) \defeq & \E^{\tilde{\gamma}_t(\cdot\vert x)} \left[ R(x_t,A_t,z_t) +\delta V_{t+1}(\phi(z_t,\tgamma_t), X_{t+1}) \vert  z_t,x_t\right].  \label{eq:Vdef}
}
   \end{itemize}
   
Then, an equilibrium strategy is defined as 
\eq{
\tilde{\sigma}_t(a_t\vert z_{1:t},x_{1:t}) = \tilde{\gamma}_t(a_t\vert x_t), \label{eq:sigma_fh}
} 
where $\tilde{\gamma}_t = \theta[z_t]$. 

In the following theorem, we show that the strategy thus constructed is an MFE of the game.

%%%%%%%%%%%%
\begin{theorem}
\label{Thm:Main}
A strategy $(\tsigma)$ constructed from the above methodology is an MFE of the game i.e. $\forall t\in[T], h_t \in \cH_t, {\sigma}$,
\eq{
&\E^{\tsigma}[\sum_{n=t}^T \delta^{n-t} R(X_n,A_n,Z_n)\vert z_{1:t},x_{1:t} ] \nn\\
&\geq \E^{{\sigma}}[\sum_{n=t}^T \delta^{n-t} R(X_n,A_n,Z_n)\vert z_{1:t},x_{1:t} ], \;\; \;\;   \label{eq:prop}
} 
and $z_{n+1} = \phi(z_n,\tsigma(\cdot\vert z_n,\cdot)) \forall n\in[T]$ such that $n\geq t$.
\end{theorem}
\begin{IEEEproof}
Please see Appendix~\ref{app:A}.
\end{IEEEproof}

The intuition for~\eqref{eq:m_FP} is as follows. It notes that at equilibrium, the update of the mean field $z_t$ is defined by the equilibrium strategies of the players, and no user has an incentive to unilaterally deviate in its action. Thus the equilibrium strategy maximizes a user's utility-to-go when the mean field is updated by the same equilibrium policy, which explains the occurrence of $\tilde{\gamma}$ at both left and right side of~\eqref{eq:m_FP}. 

In the following, we show that every MFE can be found using the above backward recursion.
\subsection{Converse}
\begin{theorem}[Converse]
	\label{thm:2}
	Let $\tsigma$ be an MFE of the mean field game. Then there exists an equilibrium generating function $\theta$
 that satisfies \eqref{eq:m_FP} in backward recursion such that  $\tsigma$ is defined using $\theta$.
\end{theorem}
\begin{IEEEproof}
	Please see Appendix~\ref{app:C}.
\end{IEEEproof}

In the following we consider the infinite horizon game $\mathbb{G}_{\infty}$ and provide a similar methodology as before to compute its MFE.

\section{Methodology for the infinite horizon problem $\mathbb{G}_{\infty}$} \label{sec:fhbr}
\label{sec:ih}
In this section, we consider the infinite-horizon problem $\mathbb{G}_{\infty}$, for which we assume the reward function $R$ to be absolutely bounded.

We define an equilibrium generating function $\theta:\cZ\to\{\cX\to\mathcal{P}(\cA) \}$, where for each $z_t $, we generate $\tgamma_t = \theta[z_t]$. In addition, we generate a reward-to-go function $V:\cZ\times\cX\to\mathbb{R}$.
These quantities are generated through a fixed-point equation as follows.

For all $z,$ set $\tilde{\gamma} = \theta[z]$. Then $(\tilde{\gamma},V)$ are solution of the following fixed-point equation\footnote{We discuss the existence of solution of this fixed-point equation in Section~\ref{sec:exists}.}, $\forall z\in\cZ,x\in \cX$,

  \eq{
 \tilde{\gamma}(\cdot\vert x) &\in \arg\max_{\gamma(\cdot\vert x)} \E^{\gamma(\cdot\vert x)} \left[ R(x,A,z) +\delta V(\phi(z,\tgamma), X^{'}) \vert  z,x\right] ,\label{eq:m_FP_ih} \\
 V(z,x) &=\  \E^{\tilde{\gamma}(\cdot\vert x)} \left[ R(x,A,z) +\delta V (\phi(z,\tgamma), X^{'}) \vert  z,x\right]. \label{eq:m_FP_ih2} 
  }
  
 where expectation in \eqref{eq:m_FP_ih} is with respect to random variable $(X^{[1:N-1]},A,X^{\prime})$ through the measure
$z(x^{[1:N-1]})\prod_{i=1}^{N-1}\gamma(a^i\vert x^i)Q_x(x^{'}\vert z,x^{[1:N-1]},x,a^{[1:N-1]},a)$.

Then an equilibrium strategy is defined as 
\eq{
\tilde{\sigma}(a_t\vert z_{1:t},x_{1:t}) = \tilde{\gamma}(a_t\vert x_t), \label{eq:sigma_ih}
}
where $\tilde{\gamma} = \theta[z_t]$.

The following theorem shows that the strategy thus constructed is an MFE of the game.

%%%%%%%%%%%%
\begin{theorem}
\label{thih}
A strategy $(\tsigma)$ constructed from the above methodology is an MFE of the game i.e. $\forall t, h_t \in \cH_t, {\sigma}$,

\eq{
&\E^{\tsigma}[\sum_{n=t}^\infty \delta^{n-t} R(X_n,A_n,Z_n)\vert z_{1:t},x_{1:t} ] \nn\\
&\geq \E^{{\sigma}}[\sum_{n=t}^\infty \delta^{n-t} R(X_n,A_n,Z_n)\vert z_{1:t},x_{1:t} ], \;\; \;\;   \label{eq:prop_ih}
} 
and $z_{n+1} = \phi(z_n,\tsigma(\cdot\vert z_n,\cdot)) \forall n\in[T]$ such that $n\geq t$.
\end{theorem}
\begin{IEEEproof}
Please see Appendix~\ref{app:D}.
\end{IEEEproof}

In the following, we show that every mean field equilibria can be found using the above backward recursion.
\subsection{Converse}
\begin{theorem}[Converse]
	\label{thm:2ih}
	Let $\tsigma$ be an MFE of the mean field game. Then there exists an equilibrium generating function $\theta$
 that satisfies \eqref{eq:m_FP_ih}-\eqref{eq:m_FP_ih2} in backward recursion such that  $\tsigma$ is defined using $\theta$.
\end{theorem}
\begin{IEEEproof}
	Please see Appendix~\ref{app:idih}.
\end{IEEEproof}

\section{Existence of per stage fixed-point equation}
\label{sec:exists}
In this section, we discuss sufficient conditions for the existence of a solution of the fixed-point equations~\eqref{eq:m_FP} and~\eqref{eq:m_FP_ih}-\eqref{eq:m_FP_ih2}.

\begin{theorem}
Under assumption~(A1), for every $t$ there exists solution of the fixed-point equations~\eqref{eq:m_FP}, and for \eqref{eq:m_FP_ih}-\eqref{eq:m_FP_ih2}.
\end{theorem}
\begin{IEEEproof}
Under the assumption (A1), it was shown in~Theorem~1 that there exists an MFE of both the finite and infinite horizon games. Furthermore, Theorem~\ref{thm:2} and Theorem~\ref{thm:2ih} show that all MFE can be found using backward recursion for the finite and infinite horizon problems respectively. This proves that under (A1), for every $t$, there exists a solution of~\eqref{eq:m_FP}, and for~\eqref{eq:m_FP_ih}-\eqref{eq:m_FP_ih2}.
\end{IEEEproof}

\section{Conclusion}
\label{sec:Concl}
In this paper, we define both finite and infinite horizon, large population dynamic game where each player is affected by others through a correlated mean-field population state. We prove the existence of MFG under appropriate sufficient conditions. We present a novel backward recursive methodology to compute Mean-field equilibria (MFE) for such games, where each player's strategy depends on its current private type and current correlated mean-field population state. We also prove the existence of each fixed-point equation $t$. This new framework opens door to studying many new applications where players have correlated types.

\begin{appendices}
\section{}
We first note that as mentioned before, the proof in this Appendix is adapted from the proof of MFE in~\cite{DeLaRa19}. As shown in~\cite{DeLaRa19}, we first note that a (mixed) strategy is a measurable
function $\sigma:\cZ\times\cX\to\calP(\calA)$, that associates to each
state $x\in\cX$ and each time $t$ a probability measure
 on the set of possible actions. We also
denote by $\pi^{\sigma}(x_t,a_t)$ the probability that, at time $t$, a player
in state $x_t$ takes the action $a_t$, under strategy $\sigma$. For all
$t$ and all $x\in\cX$, we have $\sum_{a\in\cA}\pi^{\sigma}(x,a)=1$.
The set of all possible strategies is denoted by $\strat$.

% We say that a strategy is
% {\it pure} if, for all state $i$ and all $t\in\mathbb{R}$, there
% exists an action $a\in\calA$ such that $\pi_{i,a}(t)=1$ and
% $\pi_{i,a^\prime}(t)=0$ for all $a^\prime\neq a$.

The set $\strat$ is a bounded subset of the Hilbert space of the
functions $\cZ\times\cX \to\cP(\cA)$ equipped with the inner product the
exponentially weighted inner product :
$<f,g>=\sum_{t=0}^\infty \delta^t f_tg_t$. This shows that
$\strat$ is weakly compact, where the weak topology is defined as
follows: a sequence of policy $\sigma^n$ converges to a policy
$\sigma$ if for any bounded function $g$:
\begin{align*}
  \lim_{n\to\infty}\sum_{t=0}^\infty \delta^t \sigma^n_tg_t = \sum_{t=0}^\infty\delta^t
  \sigma_tg_tdt. 
\end{align*}
\label{app:A0}

\subsubsection{Proof of Theorem~\ref{thm:mfe-existence}}
\begin{IEEEproof}

\begin{itemize}

\item Let \eq{
W(z_{1:T},\sigma_{1:T}) = \E^{{\sigma_{1:T}}}[\sum_{n=1}^T \delta^{n-t} R(X_n,A_n,z_n) ] \label{eq:Wdef}
}
\item A policy $\tsigma$ is optimal for $z := z_{1:t}$ if for all $ t\in[T], \sigma_{t:T}, x_{1:t}$,
\eq{
&\E^{\tsigma_{t:T}}[\sum_{n=t}^T \delta^{n-t} R(X_n,A_n,Z_n)\vert z_{1:t},x_{1:t} ] \geq \nn\\
&\E^{{\sigma_{t:T}}}[\sum_{n=t}^T \delta^{n-t} R(X_n,A_n,Z_n)\vert z_{1:t},x_{1:t} ], \;\; \;\;   \label{eq:seqeq2b}
} 
\item Define the above backward optimization equation $\psi: \cS_z\to2^{\cS_{\sigma}}$ as
 \eq{
\psi(z) :=\{\sigma\in S_{\sigma} : \sigma \text{ is optimal given } z\}. \label{eq:seqeq3b}
}
\item Conversely, define a forward mapping $\Lambda: \cS_{\sigma} \to \cS_{z} $ as follows: given $\sigma \in S_{\sigma}, z = \Lambda(\sigma)$, is constructed recursively as

\eq{
&z_{t+1}(x^1_{t+1},x_{t+1}^2,\ldots, x_{t+1}^N) = \sum_{x_t^{[1:N]},a_t^{[1:N]}}z_t(x^1_{t},x_{t}^2,\ldots, x_{t}^N)\times\nn\\
& Q_x(x^1_{t+1},x_{t+1}^2,\ldots, x_{t+1}^N\vert z_t,x^{[1:N]}_{t}, a_t^{[1:N]})\prod_{i=1}^N\sigma_t(a^i_t\vert z_t,x^i_t)
}

Define $\Omega:\mathcal{S}_z\to2^\mathcal{S}_z$ as the best
response to a population distribution $z$ i.e.
\eq{
\Omega(z) = \Lambda(\psi(z))
}

\textbf{Definition of $\Omega(z)$} -- Since $W$ is
continuous in $\sigma$(which is implied by the continuity of $\Lambda(\sigma)$ ). This shows that there exists $\sigma$
that attains the maximum in
Equation~\eqref{eq:seqeq3b}, which shows that $\Omega(z)$ is well
defined and non-empty.
\\

\textbf{Compactness of $\Omega(z)$} -- Let us consider the following
optimization problem:

\begin{align}
  \label{eq:Z1}
  &\min_{\bbx,\bbz}
    \sum_{t=0}^{T}{\left(\sum_{x_t,a_t} \delta^{t}\pi_t(x_t,a_t)R(z_t,x_t,a_t) \right)}\\
  & \text{such that $\pi_t$ satisfies }\nn\\
    &\left\{
    \begin{array}{ll}
      \sum_{a_t}\pi_t(x_t,a_t)=x_t &\forall x_{t}\in\calE,\\
      \pi_t(x_t,a_t)\geq 0, &\forall x_{t}\in\calE, \forall a_t\in\mathcal{A},\\
      x_{t+1}=\sum_{x_t,a_t}\pi_t(x_t,a_t)Q_x(x_{t+1}\vert z_t,x_t,a_t) &\forall x_{t+1}
    \end{array}
                                                            \right.
                                                            \label{eq:Z2}
\end{align}
The above problem is a linear problem, which implies that the set of
optimal solutions is convex and compact. Let us show that the set of
optimal solution of the optimization problem~\eqref{eq:Z1} is
$\Omega(z_t)$. To show this, let us remark that  the constraints \eqref{eq:Wdef} are
equivalent to the constraints \eqref{eq:Z2} by replacing the variables
$x_t \sigma_t(a_t|z_t,a_t)$ by $\pi(x_t,a_t)$. Then, the constraint
$\sigma_t\in\strat$ of \eqref{eq:Wdef}, that corresponds to
$\sigma_t\in\calP(\calA)$, is replaced with $\pi_t\ge0$ and
$\sum_{a_t} \pi_t(x_t,a_t)=x_t$.\\

\textbf{Upper-semi continuity of $\Omega$}. To prove that $\Omega$ is
upper-semi continuous, let us show that the graph of
$z\mapsto\Omega(z)$ is closed. Let $z_n\in\cS_z$ and
$\bbx_n\in \Omega(z_n)$ be two sequences such that
$\lim_{n\to\infty} z_n= z_\infty$ and
$\lim_{n\to\infty} \bbx_n= \bbx_\infty$. We want to show that
$\bbx_\infty\in \Omega(z_\infty)$.

As $W$ is continuous, for all $\bbx_n\in\Omega(z_n)$, there exists a
strategy $\sigma_n$ that minimizes $W(z_n,\sigma_n)$ and such that
$\bbx_n=\Lambda(\sigma_n,z_n)$. As the set $\strat$ is weakly compact,
this sequence of strategies has a sub sequence that converges weakly to
a strategy $\sigma_*$. Moreover, we have:
\begin{itemize}
\item As $W$ is continuous, $\sigma_*$ minimizes
  $W(\pi,z_\infty)$. This shows that
  $\bbx^{\sigma_*}\in\Omega(z_\infty)$.
\item The solution of \eqref{eq:Wdef} is continuous in $\sigma$ and $z$,
  which shows that $\bbx_\infty=\bbx^{\sigma_*,z_\infty}$.
\end{itemize}
Combining these two facts shows that $\bbx_\infty\in\Omega(z_\infty)$
which implies that the graph of $\Omega$ is closed. 

Since for all $z\in\mathcal{S}_z$, $\Omega(z)$ is
well defined and non empty (since the minimum is attained in~\ref{eq:seqeq2}), is
convex and compact.  Moreover, the function
$\Omega(\cdot)$ is upper-semi-continuous. As $\mathcal{S}_z$ is compact
\cite[Prop. 11.11]{border1989fixed}, this shows that $\Omega(\cdot)$
satisfies the conditions of the fixed point theorem given in
\cite[Theorem 8.6]{granas2013fixed} and therefore has a fixed point
$z^*$. By the definition of $\Omega$, this implies that there exists
a strategy $\sigma$ that is a best-response to $z$, which
implies that $\sigma$ is a mean field equilibrium.
%\qed

Similar arguments are used for the infinite horizon game.

\end{itemize}

\end{IEEEproof}
 \section{}
\label{app:A}
\begin{IEEEproof}
We prove Theorem~\ref{Thm:Main} using induction and the results in Lemma~\ref{lemma:2}, and \ref{lemma:1} proved in \ref{app:B}. Let $\tsigma$ be the strategies computed by the methodology in Section~III and let $z_{n+1} = \phi(z_n,\tsigma(\cdot\vert z_n,\cdot)) \forall n\in[T]$.

\seqn{
For base case at $t=T$, $\forall i\in [N], x_{1:T}, \sigma$
\eq{
\E^{\tsigma_{T}}\big\{  R(X_T,A_T,Z_T) \big\lvert z_{1:T}, x_{1:T} \big\}
&=
V_T(z_T, x_T)  \label{eq:T2a}\\
&\geq \E^{\sigma_{T}} \big\{ R(X_T,A_T,Z_T) \big\lvert z_{1:T}, x_{1:T} \big\},  \label{eq:T2}
}
}
where \eqref{eq:T2a} follows from Lemma~\ref{lemma:1} and \eqref{eq:T2} follows from Lemma~\ref{lemma:2} in \ref{app:B}.

Let the induction hypothesis be that for $t+1$, $\forall i\in [N], z_{1:t+1}, x_{1:t+1} \in (\cX)^{t+1}, \sigma$,
\seqn{
\eq{
 \E^{\tsigma_{t+1:T} } \big\{ \sum_{n=t+1}^T \delta^{n-t-1}R(X_n,A_n,Z_n) \big\lvert z_{1:t+1}, x_{1:t+1} \big\} \\
 \geq%\nn\\
  \E^{\sigma_{t+1:T} } \big\{ \sum_{n=t+1}^T \delta^{n-t-1} R(X_n,A_n,Z_n) \big\lvert  z_{1:t+1}, x_{1:t+1} \big\}. \label{eq:PropIndHyp}
}
}
\seqn{
Then $\forall i\in [N], x_{1:t}, \sigma$, we have
\eq{
&\E^{\tsigma_{t:T} } \big\{ \sum_{n=t}^T \delta^{n-t-1}R(X_n,A_n,Z_n) \big\lvert z_{1:t+1}, x_{1:t} \big\} \nonumber \\
&= V_t(z_{t}, x_t)\label{eq:T1}\\
&\geq \E^{\sigma_t} \big\{ R(X_t,A_t,Z_t) + \delta V_{t+1} (z_{t+1}, X_{t+1}) \big\lvert z_{1:t+1}, x_{1:t} \big\}  \label{eq:T3}\\
&= \E^{\sigma_t } \big\{ R(X_t,A_t,Z_t) +\nn\\
&\delta \E^{\tsigma_{t+1:T}} \big\{ \sum_{n=t+1}^T \delta^{n-t-1}R(X_n,A_n,Z_n) \big\lvert  z_{1:t+1}, x_{1:t},X_{t+1} \big\}  \big\vert z_{1:t+1}, x_{1:t} \big\}  \label{eq:T3b}\\
&\geq \E^{\sigma_t } \big\{ R(X_t,A_t,Z_t) +  \nn\\
&\delta\E^{\sigma_{t+1:T} } \big\{ \sum_{n=t+1}^T \delta^{n-t-1}R(X_n,A_n,Z_n) \big\lvert z_{1:t+1}, x_{1:t},X_{t+1}\big\} \big\vert z_{1:t+1}, x_{1:t} \big\}  \label{eq:T4} \\
&= \E^{\sigma_t } \big\{ R(X_t,A_t,Z_t) +   \nn\\
&\delta\E^{\sigma_{t:T} }  \big\{ \sum_{n=t+1}^T \delta^{n-t-1}R(X_n,A_n,Z_n) \big\lvert z_{1:t+1}, x_{1:t},X_{t+1}\big\} \big\vert z_{1:t}, x_{1:t} \big\}  
\label{eq:T5}\\
&=\E^{\sigma_{t:T} } \big\{ \sum_{n=t}^T \delta^{n-t}R(X_n,A_n,Z_n) \big\lvert z_{1:t},  x_{1:t} \big\}  \label{eq:T6},
}
}
where \eqref{eq:T1} follows from Lemma~\ref{lemma:1}, \eqref{eq:T3} follows from Lemma~\ref{lemma:2}, \eqref{eq:T3b} follows from Lemma~\ref{lemma:1}, \eqref{eq:T4} follows from induction hypothesis in \eqref{eq:PropIndHyp} and \eqref{eq:T5} follows since the random variables involved in the right conditional expectation do not depend on strategies $\sigma_t$.
\end{IEEEproof}

\section{}
\label{app:B}
\begin{lemma}
\label{lemma:2}
Let $\tsigma$ be the strategies computed by the methodology in Section~III and let $z_{n+1} = \phi(z_n,\tsigma(\cdot\vert z_n,\cdot)) \forall n=1\ldots t$.
Then $\forall t\in [T], i\in [N], x_{1:t}, \sigma_t$
\eq{
V_t(z_t, x_t) \geq \E^{\sigma_t} \big\{ R(X_t,A_t,Z_t) + \delta V_{t+1} (z_{t+1}, X_{t+1}) \big\lvert  z_{1:t}, x_{1:t} \big\}.\label{eq:lemma2}
}
\end{lemma}

\begin{IEEEproof}
We prove this lemma by contradiction.

 Suppose the claim is not true for $t$. This implies $\exists i, \hat{\sigma}_t, \hat{z}_{1:t},\hat{x}_{1:t}$ such that
\eq{
\E^{\hat{\sigma}_t} \big\{ R(X_t,A_t,Z_t) +  \delta V_{t+1} (Z_{t+1}, X_{t+1}) \big\lvert \hat{z}_{1:T},\hat{x}_{1:t} \big\} 
> V_t(\hat{z}_t, \hat{x}_{t}).\label{eq:E8}
}
We will show that this leads to a contradiction.
Construct 
\begin{equation}
\hat{\gamma}_t(a_t\vert x_t) = \lb{\hat{\sigma}_t(a_t\vert \hat{z}_{1:t},\hat{x}_{1:t}) \;\;\;\;\; x_t = \hat{x}_t \\ \text{arbitrary} \;\;\;\;\;\;\;\;\;\;\;\;\;\; \text{otherwise.}  }
\end{equation}

Then for $ \hat{x}_{1:t}$, we have
\seqn{
\eq{
&V_t( \hat{z}_t, \hat{x}_t) 
\nn \\
&= \max_{\gamma_t(\cdot\vert \hat{x}_t)} \E^{\gamma_t(\cdot\vert \hat{x}_t)} \big\{ R(\hat{x}_t,A_t,\hz_t) + \delta V_{t+1} (\phi(\hz_t,\tgamma_t), X_{t+1}) \big\lvert  \hat{z}_{t}, \hat{x}_{t} \big\}, \label{eq:E11}\\
&\geq\E^{\hat{\gamma}_t(\cdot\vert \hat{x}_t)} \big\{ R(x_t,A_t,\hz_t) + \delta V_{t+1} (\phi(\hz_t,\tgamma_t), {X}_{t+1}) \big\lvert  \hat{z}_t,\hat{x}_{t} \big\}   
\\ \nn 
&=\sum_{x_t^{[1:N-1]},a_t,x_{t+1}}   \big\{ R(\hat{x}_t,a_t,\hz_t) + \delta V_{t+1} (\phi(\hz_t,\tgamma_t), x_{t+1})\big\}
\hat{\gamma}_t(a_t\vert \hat{x}_t)z_t(x_t^{[1:N-1]})\nn\\
&\prod_{i=1}^{N-1}\hat{\gamma}_t(a_t^i\vert x_t^i)Q_x(x_{t+1}\vert z_t,x_t^{[1:N-1]},\hat{x}_t,a^{[1:N-1]}_t,a_t)  
\\ \nn 
&= \sum_{x_t^{[1:N-1]},a_t,x_{t+1}}  \big\{ R(\hat{x}_t,a_t,\hz_t) + \delta V_{t+1} (\phi(\hz_t,\tgamma_t), x_{t+1})\big\}
\hat{\sigma}_t(a_t\vert {\hz}_{1:t} ,\hat{x}_{1:t}) z_t(x_t^{[1:N-1]})\nn\\
&\prod_{i=1}^{N-1}\hat{\gamma}_t(a_t^i\vert x_t^i)Q_x(x_{t+1}\vert z_t,x_t^{[1:N-1]},\hat{x}_t,a^{[1:N-1]}_t,a_t)\label{eq:E9}\\
&= \E^{\hat{\sigma}_t } \big\{ R(\hat{x}_t,a_t,\hz_t)+ \delta V_{t+1} (\phi(\hz_t,\tgamma_t), X_{t+1}) \big\lvert {\hz}_{1:t},  \hat{x}_{1:t} \big\}  \\
&> V_t(\hz_t, \hat{x}_{t}), \label{eq:E10}
}
where \eqref{eq:E11} follows from definition of $V_t$ in \eqref{eq:Vdef}, \eqref{eq:E9} follows from definition of $\hat{\gamma}_t$ and \eqref{eq:E10} follows from \eqref{eq:E8}. However this leads to a contradiction.
}
\end{IEEEproof}

\begin{lemma}
\label{lemma:1}
Let $\tsigma$ be the strategies computed by the methodology in Section~III and let $z_{n+1} = \phi(z_n,\tsigma_n(\cdot\vert z_n,\cdot)) \forall n\in[T]$.
Then $\forall i\in [N], t\in [T], x_{1:t}$,
\begin{gather}
V_t(z_{t}, x_t) =
%\\ \hs{-0.2cm} 
\E^{\tsigma_{t:T}} \big\{ \sum_{n=t}^T \delta^{n-t}R(X_n,A_n,Z_n) \big\lvert  z_{1:t}, x_{1:t} \big\} .
\end{gather} 
\end{lemma}

\begin{IEEEproof}
%This lemma is in the same spirit as the following statement: ``For a controlled Markov process, if Markov policies are played, then the resulting process is a Markov process, where reward-to-go at any time can be denoted by a function of the current state."
%
\seqn{
We prove the lemma by induction. For $t=T$,
\eq{
 \E^{\tsigma_{T} } \big\{  R(X_T,A_T,Z_T) \big\lvert z_{1:T},  x_{1:T} \big\}
 &= \sum_{a_T} R(x_T,a_T,z_T) \tsigma_{T}(a_T\vert z_{T},x_{T}) \\
 &= V_T(z_{T}, x_T) \label{eq:C1},
}
}
where \eqref{eq:C1} follows from the definition of $V_t$ in \eqref{eq:Vdef}.
Suppose the claim is true for $t+1$, i.e., $\forall i\in [N], t\in [T],  x_{1:t+1}$
\begin{gather}
V_{t+1}(z_{t+1}, x_{t+1}) = \E^{\tsigma_{t+1:T}}
%\\
\big\{ \sum_{n=t+1}^T \delta^{n-t-1}R(X_n,A_n,Z_n) \big\lvert z_{1:t+1}, x_{1:t+1} \big\} 
\label{eq:CIndHyp}.
\end{gather}
Then $\forall i\in [N], t\in [T], x_{1:t}$, we have
\seqn{
\eq{
&\E^{\tsigma_{t:T} } \big\{ \sum_{n=t}^T \delta^{n-t} R(X_n,A_n,Z_n) \big\lvert  z_{1:t}, x_{1:t} \big\} 
\nonumber 
\\
&=  \E^{\tsigma_{t:T}} \big\{R(X_t,A_t,Z_t)  
\nonumber \\ 
&+\delta \E^{\tsigma_{t:T} }  \big\{ \sum_{n=t+1}^T \delta^{n-t-1}R(X_n,A_n,Z_n)\big\lvert z_{1:t+1}, x_{1:t},X_{t+1}\big\} \big\lvert z_{1:t},  x_{1:t} \big\} \label{eq:C2}
\\
&=  \E^{\tsigma_{t:T}} \big\{R(X_t,A_t,Z_t) 
\nonumber 
\\
&  +\delta\E^{\tsigma_{t+1:T}}\big\{ \sum_{n=t+1}^T \delta^{n-t-1}R(X_n,A_n,Z_n)\big\lvert z_{1:t+1}, x_{1:t},X_{t+1}\big\} \big\lvert z_{1:t}, x_{1:t} \big\} \label{eq:C3}
\\
&=  \E^{\tsigma_{t:T}} \big\{R(X_t,A_t,z_t) +  \delta V_{t+1}(z_{t+1}, X_{t+1}) \big\lvert  z_{1:t}, x_{1:t} \big\} 
\label{eq:C4}
\\
&=  \E^{\tsigma_{T}} \big\{R(X_t,A_t,z_t) +  \delta V_{t+1}(z_{t+1}, X_{t+1}) \big\lvert  z_{1:t}, x_{1:t} \big\} 
\label{eq:C5}
\\
&=V_{t}(z_t, x_t) \label{eq:C6},
}
}
\eqref{eq:C4} follows from the induction hypothesis in \eqref{eq:CIndHyp}, \eqref{eq:C5} follows because the random variables involved in expectation, $X_t,A_t,X_{t+1}$ do not depend on $\tsigma_{t+1:T}$ and \eqref{eq:C6} follows from the definition of $V_t$ in \eqref{eq:Vdef}.
\end{IEEEproof}

\section{}
\label{app:C}
\begin{IEEEproof}
We prove this by contradiction. Suppose for any equilibrium generating function $\theta$ that generates an MFE $\tsigma$ and for $z_{n+1} = \phi(z_n,\tsigma(\cdot\vert z_n,\cdot)) \forall n\in[T]$, there exists $t\in[T], i\in[N]$ such that \eqref{eq:m_FP} is not satisfied for $\theta$
%\footnote{Note that for $z_t \neq z_t $ for any $a_{1:t-1}$, $\phi$ can be arbitrarily defined without affecting the definition of $(\tsigma,\mu^*)$.}
i.e. for $\tgamma_t = \theta_t[z_t] = \tsigma_t(\cdot\vert z_t,\cdot)$,
\eq{
 \tilde{\gamma}_t(\cdot\vert x_t) \not\in \arg\max_{\gamma_t(\cdot\vert x_t)} \E^{\gamma_t(\cdot\vert x_t)} \big\{ R_t(X_t,A_t,z_t) + V_{t+1}(\phi(z_t,\tilde{\gamma}_t), X_{t+1}) \big\lvert  x_t,z_t \big\} . \label{eq:FP4}
  }
  Let $t$ be the first instance in the backward recursion when this happens. This implies $\exists\ \hat{\gamma}_t(\cdot\vert x_t)$ such that
  \eq{
  \E^{\hat{\gamma}_t(\cdot\vert x_t)} \big\{ R_t(x_t,A_t,z_t)+ V_{t+1} (\phi(z_t, \tilde{\gamma}_t), X_{t+1}) \big\lvert  z_{1:t},x_{1:t}\big\}
  \nn\\
  > \E^{\tgamma_t(\cdot\vert x_t)} \big\{ R_t(x_t,A_t) + V_{t+1} (\phi(z_t, \tilde{\gamma}_t), X_{t+1}) \big\lvert  z_{1:t},x_{1:t} \big\} \label{eq:E1}
  }
  This implies for $\hat{\sigma}_t(\cdot\vert z_t,\cdot) = \hat{\gamma}_t$,
  \eq{
  &\E^{\tsigma_{t:T}} \big\{ \sum_{n=t}^T R_n(X_n,A_n,Z_n) \big\lvert  z_{1:t}, x_{1:t} \big\}
  \nn\\
  %&= \E^{\tsigma_t} \big\{ R_t(X_t,A_t,Z_t) + \E^{\tsigma_{t:T}}  \big\{ \sum_{n=t+1}^T R_n(X_n,A_n,Z_n) \big\lvert z_{1:t},Z_{t+1}, x_{1:t},X_{t+1} \big\}  \big\vert z_{1:t}, x_{1:t} \big\}% \label{eq:E2a}
  &= \E^{\tsigma_t} \big\{ R_t(x_t,A_t,z_t) + \nn\\
  &\E^{\tsigma_{t+1:T} }   \big\{ \sum_{n=t+1}^T R_n(X_n,A_n,z_n) \big\lvert z_{1:t+1}, x_{1:t},X_{t+1} \big\}  \big\vert z_{1:t}, x_{1:t} \big\} \label{eq:E2}
  \\
  &=\E^{\tgamma_t(\cdot\vert x_t) } \big\{ R_t(x_t,A_t,z_t) + V_{t+1} (\phi(z_t, \tilde{\gamma}_t), X_{t+1}) \big\lvert  z_t,x_t \big\} \label{eq:E3}
  \\
  &< \E^{\hat{\sigma}_t(\cdot\vert z_t,x_t)} \big\{ R_t(x_t,A_t,z_t) + V_{t+1} (\phi(z_t, \tilde{\gamma}_t), X_{t+1}) \big\lvert  z_t,x_t \big\}\label{eq:E4}
  \\
  &= \E^{\hat{\sigma}_t } \big\{ R_t(x_t,A_t,z_t) + \nn\\
  & \E^{\tsigma_{t+1:T}}\big\{ \sum_{n=t+1}^T R_n(X_n,A_n,z_n) \big\lvert z_{1:t+1}, x_{1:t},X_{t+1}\big\} \big\vert z_{1:t}, x_{1:t} \big\}\label{eq:E5}
  \\
  &=\E^{\hat{\sigma}_t,\tsigma_{t+1:T} } \big\{ \sum_{n=t}^T R_n(X_n,A_n,z_n) \big\lvert  z_{1:t}, x_{1:t}\big\},\label{eq:E6}
  }
  where \eqref{eq:E3} follows from the definitions of $\tgamma_t$ and Lemma~\ref{lemma:1}, \eqref{eq:E4} follows from \eqref{eq:E1} and the definition of $\hat{\sigma}_t$, \eqref{eq:E5} follows from Lemma~\ref{lemma:2}. However, this leads to a contradiction since $\tsigma$ is an MFE of the game.
\end{IEEEproof}

\section{}
\label{app:D} Let $\tsigma$ be the strategies computed by the methodology in Section~IV and let $z_{n+1} = $ \\ $\phi(z_n,\tsigma_n(\cdot\vert z_n,\cdot)) \forall n\in[T]$.
	We divide the proof into two parts: first we show that the value function $V$ is at least as big as any reward-to-go function; secondly we show that under the strategy $\tsigma$, reward-to-go is $ V $. Note that $h_t := (z_{1:t},x_{1:t})$. 

\paragraph*{Part 1:}
For any $ i \in \mN $, $ \sigma $ define the following reward-to-go functions
\begin{subequations} \label{eqihr2g}
%	\begin{align} 				
\eq{
	W_t^{\sigma}(h_t) &= \mE^{\sigma} \lpr \sum_{n=t}^\infty \delta^{n-t} R(X_n,A_n,Z_n) \mid h_t \rpr\\
%	\\
	%\label{eqr2gfh}		
	W_t^{\sigma,T}(h_t) &= \mE^{\sigma} \lpr \sum_{n=t}^T \delta^{n-t} R(X_n,A_n,Z_n)
%	\\
	+  \delta^{T+1-t} V(Z_{T+1},X_{T+1}) \mid h_t \rpr.
}
%	\end{align}
\end{subequations}
Since $ \mX,\mA $ are finite sets the reward $ R$ is absolutely bounded, the reward-to-go $ W_t^{\sigma}(h_t) $ is finite $ \forall $ $ i,t,\sigma,h_t $.

For any $ i \in \mN $, $ x_{1:t}$,
\eq{ \label{eqdc}
%\begin{equation}
V\big(z_t,x_t\big) - W_t^{\sigma}(h_t)
= \Big[ V\big(z_t,x_t\big) - W_t^{\sigma,T}(h_t) \Big]
%	\\
+ \Big[ W_t^{\sigma,T}(h_t) - W_t^{\sigma}(h_t) \Big]
%\end{equation} 	
	}
Combining results from Lemmas~\ref{thmfh2} and~\ref{lemfhtoih} in \ref{app:D}, %
%8 and 9 in Appendix G of the technical report~\cite{VaSiAn17arxiv}, 
the term in the first bracket in RHS of~\eqref{eqdc} is non-negative. Using~\eqref{eqihr2g}, the term in the second bracket is
%	\eq{
\begin{gather}\label{eqdiff}	
\left( \delta^{T+1-t} \right) \mE^{\sigma} \Big\{- \sum_{n=T+1}^\infty \delta^{n-(T+1)} R(X_n,A_n,Z_n)
%	\\
+ V(Z_{T+1},X_{T+1}) \mid h_t \Big\}.
\end{gather} 	
%	}
The summation in the expression above is bounded by a convergent geometric series. Also, $ V $ is bounded. Hence the above quantity can be made arbitrarily small by choosing $ T $ appropriately large. Since the LHS of~\eqref{eqdc} does not depend on $ T $, which implies,
\begin{gather}
V\big(z_t,x_t\big) \ge W_t^{\sigma}(h_t).
\end{gather}

\paragraph*{Part 2:}
Since the strategy the equilibrium strategy $ \tsigma $ generated in~\eqref{eq:sigma_ih} is such that $\tsigma_t $ depends on $ h_t $ only through $ z_t $ and $ x_t $, the reward-to-go $ W_t^{\tsigma} $, at strategy $ \tsigma $, can be written (with abuse of notation) as
\begin{gather}
%\begin{gather}
W_t^{\tsigma}(h_t) = W_t^{\tsigma}(z_t,x_t)
%\\
= \mE^{\tsigma} \lpr \sum_{n=t}^\infty \delta^{n-t} R(X_n,A_n,Z_n) \mid z_t,x_t \rpr.
%\end{gather}
\end{gather}

For any $ x_{1:t}$,
\begin{subequations}
\eq{
	W_t^{\tsigma}(z_t,x_t)
	&= \mE^{\tsigma} \lpr R(X_t,A_t,Z_t)	+ \delta W_{t+1}^{\tsigma}
%	\\
	\big(\phi(z_t,\theta[z_t])),X_{t+1}\big)  \mid z_t,x_t \rpr\\
%\\
	V(z_t,x_t)
	&= \mE^{\tsigma} \Big\{ R(X_t,A_t,Z_t) + \delta V
%	\\
	\big(\phi(z_t,\theta[z_t])),X_{t+1}\big)  \mid z_t,x_t \Big\}.
%\end{gather}
}
\end{subequations}
Repeated application of the above for the first $ n $ time periods gives
\begin{subequations}
\eq{
%\begin{align}
	W_t^{\tsigma}(z_t,x_t)
	&= \mE^{\tsigma}\Bigg\{ \sum_{m=t}^{t+n-1} \delta^{m-t} R(X_t,A_t,Z_t)
%	\\
	+ \delta^{n}  W_{t+n}^{\tsigma}\big(Z_{t+n},X_{t+n}\big)  \mid z_t,x_t \Bigg\}
\\
	V(z_t,x_t)
	&= \mE^{\tsigma} \Bigg\{ \sum_{m=t}^{t+n-1} \delta^{m-t} R(X_t,A_t,Z_t)
%	\\
	+ \delta^{n}  V\big(Z_{t+n},X_{t+n}\big)  \mid z_t,x_t \Bigg\}.
%\end{align}
}
\end{subequations}
Taking differences results in
%\begin{subequations}
	\eq{
	W_t^{\tsigma}(z_t,x_t)  - V(z_t,x_t)
	=\delta^n \mE^{\tsigma}
%	\\
	  \lpr W_{t+n}^{\tsigma}\big(Z_{t+n},X_{t+n}\big)
	- V\big(Z_{t+n},X_{t+n}\big) \mid z_t,x_t \rpr.
	}
%\end{subequations}
Taking absolute value of both sides then using Jensen's inequality for $ f(x) = \vert x \vert $ and finally taking supremum over $ h_t $ reduces to
\eq{
&\sup_{h_t} \big\vert W_t^{\tsigma}(z_t,x_t)  - V(z_t,x_t) \big\vert \nn\\
&\le \delta^n \sup_{h_t}  \mE^{\tsigma}
%\\
 \lpr\big\vert W_{t+n}^{\tsigma}(Z_{t+n},X_{t+n})
 - V(Z_{t+n},X_{t+n}) \big\vert  \mid z_t,x_t \rpr.\
}
Now using the fact that $ W_{t+n},V$ are bounded and that we can choose $ n $ arbitrarily large, we get $ \sup_{h_t} \vert W_t^{\tsigma}(z_t,x_t)  - V(z_t,x_t) \vert = 0 $. 	

\section{}
\label{app:E}
In this section, we present three lemmas. Lemma~\ref{thmfh1} is intermediate technical results needed in the proof of Lemma~\ref{thmfh2}. Then the results in Lemma~\ref{thmfh2} and~\ref{lemfhtoih} are used in \ref{app:C} for the proof of Theorem~\ref{thih}. The proof for Lemma~\ref{thmfh1} below isn't stated as it analogous to the proof of Lemma~\ref{lemma:2} from \ref{app:B}, used in the proof of Theorem~\ref{Thm:Main} (the only difference being a non-zero terminal reward in the finite-horizon model).

Let $\tsigma$ be the strategies computed by the methodology in Section~IV and let $z_{n+1} = \phi(z_n,\tsigma_n(\cdot\vert z_n,\cdot)) \forall n\in[T]$. Define the reward-to-go $ W_t^{\sigma,T} $ for any agent $ i $ and strategy $ \sigma $  as
%\eq{
\eq{\label{eqr2gfh}
&W_t^{\sigma,T}(z_{1:t},x_{1:t}) = \nn\\
&\mE^{\sigma} \big[ \sum_{n=t}^T \delta^{n-t} R(X_n,A_n,Z_n)
%\\
+ \delta^{T+1-t} G(Z_{T+1},X_{T+1}) \mid z_{1:t},x_{1:t} \big].
}
%}
Since $ \mX,\mA $ are assumed to be finite and $ G $ absolutely bounded, the reward-to-go is finite $ \forall $ $ i,t,\sigma,x_{1:t} $.
In the following, any quantity with a $T$ in the superscript refers the finite horizon model with terminal reward $G$.

Let $V_t^T(z_t,x_t)$ be the value function for the finite time horizon problem with horizon $T$ defined in~\eqref{eq:Vdef}.

\begin{lemma}\label{thmfh1}
	For any $ t \in [T] $, $ i \in \mN $, $x_{1:t} $ and $ \sigma $,	
	\begin{equation} \label{eqintlem}
	V_t^{T}(z_t,x_t) \ge \mE^{\sigma} \big[ R(x_t,A_t,z_t)
	%	\\
	+ \delta V_{t+1}^{T}\big( \phi(z_t,\theta[z_t]) , X_{t+1} \big) \mid z_{1:t},x_{1:t} \big].
	\end{equation}
	%	}
\end{lemma}

%\begin{lemma} \label{lemcond}
%	\eq{
%	\mE^{\sigma_{t+1:T},\tsigma_{t+1:T}^{-i}} \big[ \sum_{n=t+1}^T \delta^{n-(t+1)} R(X_n,A_n,Z_n)
%	%	\\
%	+ \delta^{T+1-t} G(Z_{T+1},X_{T+1}) \mid z_{1:t},x_{1:t},z_t,x_{t+1} \big]
%	\\
%	= \mE^{\sigma_{t:T},\tsigma_{t:T}^{-i}} \big[ \sum_{n=t+1}^T \delta^{n-(t+1)} R(X_n,A_n,Z_n)
%	%	\\
%	+ \delta^{T+1-t} G(Z_{T+1},X_{T+1}) \mid z_{1:t},x_{1:t},z_t,x_{t+1} \big].
%	}
%\end{lemma}
The result below shows that the value function from the backwards recursive methodology is higher than any reward-to-go.

\begin{lemma}\label{thmfh2}
	For any $ t \in [T] $, $ i \in \mN $, $ x_{1:t} $ and $ \sigma $,	
	\begin{gather}
	V_t^{T}(z_t,x_t) \ge W_t^{\sigma,T}(z_{1:t},x_{1:t}).
	\end{gather}
\end{lemma}
\begin{IEEEproof}
We use backward induction for this. At time $ T $, using the maximization property from~\eqref{eq:m_FP} (modified with terminal reward $ G $),
\begin{subequations}
	\begin{align}
	&V_T^{T}(z_T,x_T)
	\\
	&\triangleq \mE^{\tilde{\gamma}_T^{T}(\cdot \mid x_T)} \big[ R(X_T,A_T,Z_T) + \delta G\big( \phi(z_T,\tilde{\gamma}_T^T)),X_{T+1} \big) \mid z_T,x_T \big]
	\\
	&\ge \mE^{{\gamma}_T^{T}(\cdot \mid x_T)} \big[ R(X_T,A_T,Z_T)
	+ \delta G\big( \phi(z_T,\tilde{\gamma}_T^T)) ,X_{T+1} \big) \mid z_{1:T},x_{1:T} \big]
	\\
	&= W_T^{\sigma,T}(h_T)
	\end{align}
\end{subequations}
Here the second inequality follows from~\eqref{eq:m_FP} and~\eqref{eq:Vdef} and the final equality is by definition in~\eqref{eqr2gfh}.

Assume that the result holds for all $ n \in \{t+1,\ldots,T\} $, then at time $ t $ we have
\begin{subequations}
	\begin{align}
	&V_t^{T}(z_t,x_t)
	\\
	&\ge \mE^{\sigma_t} \big[ R(X_t,A_t,Z_t)
	+ \delta V_{t+1}^{T}\big( \phi(z_t,\theta[z_t]) , X_{t+1} \big) \mid z_{1:t},x_{1:t} \big]
	\\
	&\ge \mE^{\sigma_t} \big[ R(X_t,A_t,Z_t)
	+ \delta \mE^{\sigma_{t+1:T}} \big[ \sum_{n=t+1}^T \delta^{n-(t+1)} R(X_n,A_n,Z_n)
	\\ \nonumber
	&+ \delta^{T-t} G(Z_{T+1},X_{T+1}) \mid z_{1:t},x_{1:t},X_{t+1} \big] \mid z_{1:t},x_{1:t} \big]
	\\
	&= \mE^{\sigma_{t:T}} \big[ \sum_{n=t}^T \delta^{n-t} R(X_n,A_n,Z_n)
	+ \delta^{T+1-t}G(Z_{T+1},X_{T+1}) \mid z_{1:t},x_{1:t} \big]
	\\
	&= W_t^{\sigma,T}(z_{1:t},x_{1:t})
	\end{align}
\end{subequations}
Here the first inequality follows from Lemma~\ref{thmfh1}, the second inequality from the induction hypothesis, the third equality follows since the random variables on the right hand side do not depend on $\sigma_t$, and the final equality by definition~\eqref{eqr2gfh}.
\end{IEEEproof}

The following result highlights the similarities between the fixed-point equation in infinite-horizon and the backwards recursion in the finite-horizon.

\begin{lemma}\label{lemfhtoih}
	Consider the finite horizon game with $ G \equiv V $. Then $ V_t^{T} = V$,  $ \forall $ $ i \in \mN $, $ t \in \{1,\ldots,T\} $ satisfies the backwards recursive construction stated above (adapted from \eqref{eq:m_FP} and \eqref{eq:Vdef}).

\end{lemma}	
\begin{IEEEproof}%[Proof of Lemma~\ref{lemfh}]
	Use backward induction for this. Consider the finite horizon methodology at time $ t=T $, noting that $ V_{T+1}^{T} \equiv G \equiv  V $,
	\begin{subequations} \label{eqfhT}
		\begin{align} 	
		%		\eq{
		\tilde{\gamma}_T^{T}(\cdot \mid x_T) &\in \arg\max_{\gamma_T(\cdot \mid x_T)} \!\!\! \mE^{\gamma_T(\cdot \mid x_T)} \big[ R(x_T,A_T,z_T)
		%	\\
		+ \delta V\big( \phi(z_T,\tgamma_t^T) , X_{T+1} \big) \mid z_T,x_T \big]
		%		}
		%		\eq{
		\\
		V_T^{T}(z_T,x_T) &= \mE^{\tilde{\gamma}_T^{T}(\cdot \mid x_T)} \big[ R(x_T,A_T,z_T)
		%	\\
		+ \delta V\big( \phi(z_T,\tgamma_t^T) , X_{T+1} \big) \mid z_T,x_T \big].
		%		}
		\end{align}
	\end{subequations}
	Comparing the above set of equations with~\eqref{eq:m_FP_ih}, we can see that the pair $ (V,\tilde{\gamma}) $ arising out of~\eqref{eq:m_FP_ih} satisfies the above. Now assume that $ V_n^{T} \equiv V $ for all $ n \in \{t+1,\ldots,T\} $. At time $ t $, in the finite horizon construction from~\eqref{eq:m_FP},~\eqref{eq:Vdef}, substituting $ V$ in place of $ V_{t+1}^{T} $ from the induction hypothesis, we get the same set of equations as~\eqref{eqfhT}. Thus $ V_t^{T} \equiv V $ satisfies it.
\end{IEEEproof}

\section{}
\label{app:idih}
\begin{IEEEproof}
We prove this by contradiction. Suppose for the equilibrium generating function $\theta$ that generates MFE $\tsigma$ and for $z_{n+1} = \phi(z_n,\tsigma_n(\cdot\vert z_n,\cdot)) \forall n\in[T]$, there exists $t\in[T], i\in[N],$ such that \eqref{eq:m_FP_ih}--\eqref{eq:m_FP_ih2} is not satisfied for $\theta$
%\footnote{Note that for $z_t \neq z_t $ for any $a_{1:t-1}$, $\phi$ can be arbitrarily defined without affecting the definition of $(\tsigma,\mu^*)$.}
i.e. for $\tgamma_t = \theta[z_t] = \tsigma(\cdot\vert z_t,\cdot)$,
\eq{
 \tilde{\gamma}_t \not\in \arg\max_{\gamma_t(\cdot\vert x_t)} \E^{\gamma_t(\cdot\vert x_t)} \big\{ R(X_t,A_t,Z_t) + \delta V(\phi(Z_t,\tilde{\gamma}_t), X_{t+1}) \big\lvert  x_t,z_t \big\} . \label{eq:FP4}
  }
  Let $t$ be the first instance in the backward recursion when this happens. This implies $\exists\ \hat{\gamma}_t$ such that
  \eq{
  \E^{\hat{\gamma}_t(\cdot\vert x_t)} \big\{ R(X_t,A_t,Z_t)+ \delta V(\phi(Z_t, \tilde{\gamma}_t), X_{t+1}) \big\lvert  z_{t},x_{t}\big\}
  \nn\\
  > \E^{\tgamma_t(\cdot\vert x_t)} \big\{ R(X_t,A_t,Z_t) +\delta V(\phi(Z_t, \tilde{\gamma}_t), X_{t+1}) \big\lvert  z_{t},x_{t} \big\} \label{eq:E1}
  }
  This implies for $\hat{\sigma}(\cdot\vert z_t,\cdot) = \hat{\gamma}_t$,
  \eq{
  &\E^{\tsigma} \big\{ \sum_{n=t}^{\infty} \delta^{n-t} R(X_n,A_n,Z_n) \big\lvert  z_{1:t}, x_{1:t} \big\}
  \nn\\
  %&= \E^{\tsigma_t} \big\{ R_t(X_t,A_t,Z_t) + \E^{\tsigma_{t:T}}  \big\{ \sum_{n=t+1}^T R_n(X_n,A_n,Z_n) \big\lvert z_{1:t},Z_{t+1}, x_{1:t},X_{t+1} \big\}  \big\vert z_{1:t}, x_{1:t} \big\}% \label{eq:E2a}
\\
  &= \E^{\tsigma_t} \big\{ R(X_t,A_t,Z_t) + \nn\\
  &\E^{\tsigma_{t+1:T}}   \big\{ \sum_{n=t+1}^{\infty} \delta^{n-t}R(X_n,A_n,Z_n) \big\lvert z_{1:t}, x_{1:t},X_{t+1} \big\}  \big\vert z_{1:t}, x_{1:t} \big\} \label{eq:E2}
  }
  \eq{
  &=\E^{\tgamma_t(\cdot\vert x_t)} \big\{ R(X_t,A_t,Z_t) +\delta V (\phi(Z_t, \tilde{\gamma}_t), X_{t+1}) \big\lvert  z_t,x_t \big\} \label{eq:E3}
  \\
  &< \E^{\hat{\sigma}_t(\cdot\vert z_t,x_t)} \big\{ R(X_t,A_t,Z_t) +\delta V(\phi(Z_t, \tilde{\gamma}_t), X_{t+1}) \big\lvert  z_t,x_t \big\}\label{eq:E4}
  \\
  &= \E^{\hat{\sigma}_t} \big\{ R(X_t,A_t,Z_t) + \nn\\
  & \E^{\tsigma_{t+1:T}}\big\{ \sum_{n=t+1}^{\infty} \delta^{n-t}R(X_n,A_n,Z_n) \big\lvert z_{1:t},x_{1:t},X_{t+1}\big\} \big\vert z_{1:t}, x_{1:t} \big\}\label{eq:E5}
  \\
  &=\E^{\hat{\sigma}_t,\tsigma_{t+1:T}} \big\{ \sum_{n=t}^{\infty} \delta^{n-t} R(X_n,A_n,Z_n) \big\lvert  z_{1:t}, x_{1:t} \big\},\label{eq:E6}
  }
  where \eqref{eq:E3} follows from the definitions of $\tgamma_t$ and Appendix~\ref{app:D}, \eqref{eq:E4} follows from \eqref{eq:E1} and the definition of $\hat{\sigma}_t$, \eqref{eq:E5} follows from Appendix~\ref{app:D}. However, this leads to a contradiction since $\tsigma$ is an MFE of the game.
\end{IEEEproof}

\end{appendices}

%\section*{References}
%\bibliographystyle{model4-names}
%\bibliographystyle{IEEEtran}

%%%
%
\vspace{-0.5cm}
%\input{Deepanshu_Bio}

\bibliographystyle{IEEEtran}

%\bibliography{deepanshu,library}

\begin{thebibliography}{10}
\providecommand{\url}[1]{#1}
\csname url@samestyle\endcsname
\providecommand{\newblock}{\relax}
\providecommand{\bibinfo}[2]{#2}
\providecommand{\BIBentrySTDinterwordspacing}{\spaceskip=0pt\relax}
\providecommand{\BIBentryALTinterwordstretchfactor}{4}
\providecommand{\BIBentryALTinterwordspacing}{\spaceskip=\fontdimen2\font plus
\BIBentryALTinterwordstretchfactor\fontdimen3\font minus
  \fontdimen4\font\relax}
\providecommand{\BIBforeignlanguage}[2]{{%
\expandafter\ifx\csname l@#1\endcsname\relax
\typeout{** WARNING: IEEEtran.bst: No hyphenation pattern has been}%
\typeout{** loaded for the language `#1'. Using the pattern for}%
\typeout{** the default language instead.}%
\else
\language=\csname l@#1\endcsname
\fi
#2}}
\providecommand{\BIBdecl}{\relax}
\BIBdecl

\bibitem{LaLi07}
J.-M. Lasry and P.-L. Lions, ``Mean field games,'' \emph{Japanese Journal of
  Mathematics}, vol.~2, no.~1, pp. 229--260, 2007.

\bibitem{HuMaCa06}
M.~Huang, R.~P. Malham{\'e}, and P.~E. Caines, ``Large population stochastic
  dynamic games: closed-loop mckean-vlasov systems and the nash certainty
  equivalence principle,'' \emph{Communications in Information \& Systems},
  vol.~6, no.~3, pp. 221--252, 2006.

\bibitem{La08}
J.-M. Lasry, P.-L. Lions, and O.~Gu{\'e}ant, ``Application of mean field games
  to growth theory,'' 2008.

\bibitem{GuLaLi11}
O.~Gu{\'e}ant, J.-M. Lasry, and P.-L. Lions, ``Mean field games and
  applications,'' in \emph{Paris-Princeton lectures on mathematical finance
  2010}.\hskip 1em plus 0.5em minus 0.4em\relax Springer, 2011, pp. 205--266.

\bibitem{SuMa19}
J.~Subramanian and A.~Mahajan, ``Reinforcement learning in stationary
  mean-field games,'' in \emph{International Conference on Autonomous Agents
  and Multiagent Systems (AAMAS)}, 2019.

\bibitem{HuMa16}
M.~Huang and Y.~Ma, ``Mean field stochastic games: Monotone costs and threshold
  policies,'' in \emph{2016 IEEE 55th Conference on Decision and Control
  (CDC)}.\hskip 1em plus 0.5em minus 0.4em\relax IEEE, 2016, pp. 7105--7110.

\bibitem{HUMa17}
------, ``Mean field stochastic games with binary action spaces and monotone
  costs,'' \emph{arXiv preprint arXiv:1701.06661}, 2017.

\bibitem{HuMa17cdc}
------, ``Mean field stochastic games with binary actions: Stationary threshold
  policies,'' in \emph{2017 IEEE 56th Annual Conference on Decision and Control
  (CDC)}.\hskip 1em plus 0.5em minus 0.4em\relax IEEE, 2017, pp. 27--32.

\bibitem{AdJoWe15}
S.~Adlakha, R.~Johari, and G.~Y. Weintraub, ``Equilibria of dynamic games with
  many players: Existence, approximation, and market structure,'' \emph{Journal
  of Economic Theory}, vol. 156, pp. 269--316, 2015.

\bibitem{VaSiAn16arxiv}
D.~Vasal, A.~Sinha, and A.~Anastasopoulos, ``A systematic process for
  evaluating structured perfect bayesian equilibria in dynamic games with
  asymmetric information,'' \emph{IEEE Transactions on Automatic Control},
  2018.

\bibitem{VaAn16allerton}
\BIBentryALTinterwordspacing
D.~Vasal and A.~Anastasopoulos, ``Decentralized {B}ayesian learning in dynamic
  games,'' in \emph{Allerton Conference on Communication, Control, and
  Computing}, 2016. [Online]. Available: \url{https://arxiv.org/abs/1607.06847}
\BIBentrySTDinterwordspacing

\bibitem{VaAn16cdc}
------, ``Signaling equilibria of dynamic {LQG} games with asymmetric
  information,'' in \emph{Conference on {D}ecision and {C}ontrol}, 2016.

\bibitem{OuTaTe17}
Y.~Ouyang, H.~Tavafoghi, and D.~Teneketzis, ``Dynamic games with asymmetric
  information: Common information based perfect bayesian equilibria and
  sequential decomposition,'' \emph{IEEE Transactions on Automatic Control},
  vol.~62, no.~1, pp. 222--237, 2017.

\bibitem{Ta17}
H.~T. Jahormi, ``On design and analysis of cyber-physical systems with
  strategic agents,'' Ph.D. dissertation, University of Michigan, Ann Arbor,
  2017.

\bibitem{VaBe20}
D.~Vasal and R.~Berry, ``$ alpha-$ robust equilibrium in anonymous games,''
  \emph{arXiv preprint arXiv:2005.06812}, 2020.

\bibitem{DoGaGa19}
J.~Doncel, N.~Gast, and B.~Gaujal, ``Discrete mean field games: Existence of
  equilibria and convergence,'' \emph{arXiv preprint arXiv:1909.01209}, 2019.

\bibitem{NaMaTe13}
A.~Nayyar, A.~Mahajan, and D.~Teneketzis, ``Decentralized stochastic control
  with partial history sharing: A common information approach,''
  \emph{Automatic Control, IEEE Transactions on}, vol.~58, no.~7, pp.
  1644--1658, 2013.

\bibitem{ArMa14}
J.~Arabneydi and A.~Mahajan, ``Team optimal control of coupled subsystems with
  mean-field sharing,'' in \emph{53rd IEEE Conference on Decision and
  Control}.\hskip 1em plus 0.5em minus 0.4em\relax IEEE, 2014, pp. 1669--1674.

\bibitem{KuVa86}
P.~Kumar and P.~Varaiya, ``Stochastic systems,'' 1986.

\bibitem{DeLaRa19}
\BIBentryALTinterwordspacing
F.~Delarue, D.~Lacker, and K.~Ramanan, ``From the master equation to mean field
  game limit theory: a central limit theorem,'' \emph{Electron. J. Probab.},
  vol.~24, p. 54 pp., 2019. [Online]. Available:
  \url{https://doi.org/10.1214/19-EJP298}
\BIBentrySTDinterwordspacing

\bibitem{border1989fixed}
K.~C. Border, \emph{Fixed point theorems with applications to economics and
  game theory}.\hskip 1em plus 0.5em minus 0.4em\relax Cambridge university
  press, 1989.

\bibitem{granas2013fixed}
A.~Granas and J.~Dugundji, \emph{\em Fixed point theory}.\hskip 1em plus 0.5em
  minus 0.4em\relax Springer Science \& Business Media, 2013.

\end{thebibliography}
%\end{comment}
\end{document}